\documentclass[reprint,amsmath,amssymb,aip,jcp]{revtex4-1}

\usepackage{graphicx}
\usepackage{dcolumn}
\usepackage{bm}

 \usepackage{color}

\usepackage{amssymb}
\usepackage{amsmath}
\usepackage{graphicx}
\usepackage{amsbsy}

\newcommand\beq{\begin{equation}}
\newcommand\eeq{\end{equation}}
\newcommand\beqa{\begin{eqnarray}}
\newcommand\eeqa{\end{eqnarray}}
\newcommand{\nn}{\nonumber\\}

\newcommand{\GG}{\mathcal{G}}

\begin{document}



\title{Inferring the equation of state of a metastable hard-sphere fluid from the equation of state of a hard-sphere mixture at high densities}


\author{Andr\'es Santos}
\email{andres@unex.es}
\homepage{http://www.unex.es/eweb/fisteor/andres/}
\author{Santos B. Yuste}
\email{santos@unex.es}
\homepage{http://www.unex.es/eweb/fisteor/santos/}
\affiliation{Departamento de F\'{\i}sica, Universidad de
Extremadura, Badajoz, E-06071, Spain}
\author{Mariano L\'{o}pez de Haro}
\email{malopez@servidor.unam.mx}
\homepage{http://miquiztli.cie.unam.mx/xml/tc/ft/mlh/}
\affiliation{Centro de Investigaci\'on en Energ\'{\i}a, UNAM,
Temixco, Morelos 62580, M{e}xico}

\date{\today}

\begin{abstract}
A possible approximate route to obtain the equation of state of the monodisperse hard-sphere system in the metastable fluid region from the knowledge of the equation of state of a hard-sphere mixture at high densities is discussed. The proposal is illustrated by using recent Monte Carlo simulation data for the pressure of a binary mixture. It is further {shown} to exhibit high internal consistency.
\end{abstract}

\date{\today}


\maketitle

As is well known,  simulation studies of the metastable fluid branch of the one-component hard-sphere (HS) system, especially near the putative glass transition, are extremely difficult. This is due to the natural tendency of the system to  crystal formation. On the other hand, the introduction of a certain degree of size polydispersity hinders the formation of local ordered configurations, thus providing access to disordered states. The question naturally arises as to whether it is possible to infer the equation of state of the underlying metastable pure fluid from measurements made on the polydisperse one. The aim of this work is to provide a constructive affirmative answer to that question using  recent simulation data of the equation of state of a HS binary mixture at high densities.\cite{OB11}

One of the great assets of using the distribution function approach in the theory of liquids is the ability to link the structural properties (radial distribution functions) and the thermodynamic properties (internal energy, pressure, chemical potential, \ldots) of a given system. In the case of a multicomponent additive HS mixture such a link leads to a particularly simple form of the equation of state. Let $\rho $ be the number density of the mixture  and $x_{i}=\rho _{i}/\rho $, where $\rho_i$ is the number density of
species $i$, be the mole fraction of species $i$. Also, let $\sigma_{ij}=\frac{1}{2}(\sigma _{i}+\sigma _{j})$, where the diameter of a sphere of
species $i$ is $\sigma _{ii}=\sigma _{i}$, be the distance of separation at contact between the centers of two
interacting particles, one of species $i$ and the other one of
species $j$. In terms of the packing
fraction $\eta =(\pi/6) \rho \langle \sigma^3\rangle$, where  $\langle \sigma^n\rangle\equiv \sum_{i=1}^N x_i \sigma_i^n$ and $N$ is the number of components, the compressibility factor of the
mixture $Z=p/\rho k_{B}T$ ($p$ being the pressure, $k_{B}$ the Boltzmann constant, and
$T$ the absolute temperature) is given by
\begin{equation}
Z(\eta )=1+\frac{4 \eta}{\langle \sigma^3\rangle}\sum_{i,j=1}^N
{x_{i}x_{j}{\sigma _{ij}^{3} }} g_{ij}(\sigma_{ij}^+).
\label{1}
\end{equation}
Here $g_{ij}(\sigma_{ij}^+)$ are the contact values of the radial distribution functions $g_{ij}(r)$, $r$ being the distance. Thus, one only needs to specify these contact values in order to get the equation of state of the mixture. Unfortunately, no exact analytical results are available for them although many valuable approximate expressions have been derived or proposed and they may also be determined from computer simulations. Among the analytical expressions for the contact values, one should especially mention the ones which follow from the solution
of the Percus--Yevick (PY) equation of additive HS mixtures by
Lebowitz\cite{L64} given by
\beq
g_{ij}^{\text{PY}}(\sigma_{ij}^+)=\frac{1}{1-\eta
}+\frac{3}{2}\frac{\eta }{(1-\eta )^{2}}\frac{\sigma_i\sigma_j
\langle \sigma^2 \rangle}{\sigma_{ij}\langle \sigma^3 \rangle},
\label{15PY}
\eeq
and the results obtained from the Scaled Particle
Theory (SPT),\cite{RFL59,HFL61,LHP65,MR75,R88,HC04b} namely
\beqa
g_{ij}^{\text{SPT}}(\sigma_{ij}^+)&=&\frac{1}{1-\eta
}+\frac{3}{2}\frac{\eta }{(1-\eta )^{2}}\frac{\sigma_i\sigma_j
\langle \sigma^2 \rangle}{\sigma_{ij}\langle \sigma^3 \rangle}\nn
&&+\frac{3}{4}\frac{\eta^{2}}{(1-\eta)^{3}}\left(\frac{\sigma_i\sigma_j
\langle \sigma^2 \rangle}{\sigma_{ij}\langle \sigma^3 \rangle}\right)^{2}.
\label{15SPT}
\eeqa

Since neither the PY nor the SPT lead to particularly accurate
values when compared with simulation results, Boubl\'{\i}k\cite{B70} and, independently, Grundke
and Henderson\cite{GH72} and Lee and Levesque\cite{LL73} proposed
an interpolation between the PY and the SPT contact values, yielding what we
will refer to as the BGHLL values:
\beqa
g_{ij}^{\text{BGHLL}}(\sigma_{ij}^+)&=&\frac{1}{1-\eta
}+\frac{3}{2}\frac{\eta }{(1-\eta )^{2}}\frac{\sigma_i\sigma_j
\langle \sigma^2 \rangle}{\sigma_{ij}\langle \sigma^3 \rangle}\nn
&&+\frac{1}{2}\frac{\eta^{2}}{(1-\eta)^{3}}\left(\frac{\sigma_i\sigma_j
\langle \sigma^2 \rangle}{\sigma_{ij}\langle \sigma^3 \rangle}\right)^{2}.
\label{15BGHLL}
\eeqa
These contact values, when substituted into Eq.\ (\ref{1}), lead to the widely used and rather
accurate Boubl\'{\i}k--Mansoori--Carnahan--Starling--Leland (BMCSL)
equation of state\cite{B70,MCSL71} for HS mixtures. Refinements of the BGHLL
values have been subsequently introduced, among others, by Henderson
{et al.},\cite{HMLC96,YCH96,YCH97,HC98,HC98b,HC00,HBCW98,MHC99,CCHW00}, Matyushov and
Ladanyi,\cite{ML97} and Barrio and Solana\cite{BS00} to cope with the so-called colloidal limit of binary
mixtures. On a different path, but also having to do with the
colloidal limit, Viduna and Smith\cite{VS02a,VS02b} have proposed a method
to obtain contact values of the radial distribution functions of HS mixtures from a given equation of state.

In the search for a more general framework able to deal with arbitrary spatial dimension $d$ and number of components $N$, in our previous work\cite{SYH99,SYH02,SYH05,HYS06,HYS08,SYHAH09} we have introduced a methodology that provides the contact values of the radial distribution functions of $d$-dimensional multicomponent hard-core mixtures. In our approach, we have taken into account exact results that these contact values should satisfy, namely the limit in which one of the species, say $i$, is made of point particles, {i.e.}, $\sigma _{i}\rightarrow 0$, the limit when all the species have the same size, $\{\sigma _{k}\}\rightarrow \sigma $, the limit stemming from a binary mixture in which one of the species (say $i=1$) is much larger than the other one ({i.e.}, $\sigma _{1}/\sigma _{2}\rightarrow \infty $), but occupies a negligible volume ({i.e.}, $x_{1}(\sigma _{1}/\sigma_{2})^{d}\rightarrow 0$), and the limit when
the mixture is in contact with a hard wall. We have further assumed that, at a given packing fraction $\eta$, the dependence of $g_{ij}(\sigma_{ij}^+)$ on the parameters $\{\sigma _{k}\}$ and $\{x_{k}\}$ takes place \textit{only} through the scaled quantity
\begin{equation}
z_{ij}\equiv \frac{\sigma _{i}\sigma_{j}}{\sigma
_{ij}}\frac{\langle \sigma^{d-1} \rangle}{\langle \sigma^d \rangle}.
\label{zij}
\end{equation}
That is, we have specifically assumed that
\begin{equation}
g_{ij}(\sigma_{ij}^+)=\GG(\eta,z_{ij}),
\label{gij}
\end{equation}
where, for a given $d$, the function $\GG(\eta,z)$ is \textit{universal} in the sense that it is a common function for all the pairs $(i,j)$, regardless of the {composition, size distribution, and} number of components of the mixture. This `universality' assumption has been successfully tested in different contexts for HS mixtures in the stable fluid region (see Refs.\
\onlinecite{SYH02,SYH05,HYS06,HYS08,SYHAH09} for details).

As can be seen in Eqs.\ \eqref{15PY}--\eqref{15BGHLL}, the corresponding contact values are consistent with the assumption \eqref{gij}, where $z_{ij}$ is given by Eq.\ \eqref{zij} with $d=3$ and $\GG(\eta,z_{ij})$ has a linear (PY) or quadratic (SPT and BGHLL) dependence on $z_{ij}$. In a more general and flexible approach, the contact values are given by\cite{SYH05,HYS06,HYS08}
\beq
g_{ij}(\sigma_{ij}^+)=G_0(\eta)+G_1(\eta) z_{ij}+G_2(\eta) z_{ij}^2+G_3(\eta) z_{ij}^3,
\label{5}
\eeq
where
\beq
G_0(\eta)=\frac{1}{1-\eta},\quad G_1(\eta)=\frac{3 \eta}{2\left(1-\eta\right)^2},
\label{6}
\eeq
\beq G_2(\eta)=\frac{3
\eta^2}{4 \left(1-\eta\right)^3}-\frac{2-\eta}{1-\eta}G_3(\eta),
\label{n7b}
\eeq
\beq
G_3(\eta)=(1-\eta)\left[g_{p}^{\text{SPT}}(\eta)-g_{p}(\eta)\right],
\label{n6b}
\eeq
with
\beq
g_{p}^{\text{SPT}}(\eta)=\frac{1-\eta/2+\eta^2/4}{(1-\eta)^3}
\label{SPT}
\eeq
being the contact value of the pure HS fluid in the SPT\cite{RFL59} and $g_{p}(\eta)$ being the contact value of the pure HS fluid, which remains unspecified until one chooses a particular equation of state for this latter system.
Note that, if one chooses $g_{p}(\eta)=g_{p}^{\text{SPT}}(\eta)$, Eq.\ \eqref{5} reduces to Eq.\ \eqref{15SPT}. On the other hand, Eq.\ \eqref{5} accommodates any desired function $g_{p}(\eta)$, thus extending the contact value of the pure fluid to the contact values of a mixture with arbitrary composition, number of components, and size distribution.
Inserting Eq.\ \eqref{5} into Eq.\ \eqref{1} one obtains
\beqa
Z(\eta)&=&\frac{1}{1-\eta}+\left(\frac{\langle \sigma\rangle \langle \sigma^2\rangle}{\langle \sigma^3\rangle}-\frac{\langle \sigma^2\rangle^3}{\langle \sigma^3\rangle^2}\right)\frac{3\eta}{(1-\eta)^2}\nn
&&+\frac{\langle \sigma^2\rangle^3}{\langle \sigma^3\rangle^2}\left[Z_p(\eta)-\frac{1}{1-\eta}\right],
\label{Ze3}
\eeqa
where $Z_p(\eta)=1+4\eta g_p(\eta)$ is the compressibility factor of the pure HS fluid.
It can be checked that insertion of the PY, SPT, and Carnahan--Starling\cite{CS69} (CS) expressions for $Z_p(\eta)$ into Eq.\ \eqref{Ze3} yields their respective PY, SPT, and BMCSL extensions to mixtures.

Very recently, Odriozola and Berthier,\cite{OB11} in the search for a thermodynamic signature of the glass transition in HSs, have reported Monte Carlo {(MC)} simulation values of the radial distribution functions and the compressibility factor for a HS binary mixture up to a packing fraction of $\eta \simeq 0.63$. In view of these new results, one can now  investigate whether  (i) our proposal (Eq.\ \eqref{Ze3}), used in  a different way, allows us to infer the equation of state of the pure HS fluid in the metastable region, and (ii) the related proposal (Eq.\ \eqref{5}) still holds for high densities.

Let us begin by noting that Eq.\ \eqref{Ze3} expresses  $Z(\eta)$ in terms of {$Z_p(\eta)$ and known functions of $\eta$}. In turn, one may invert Eq.\ \eqref{Ze3} and express  $Z_p(\eta)$ in terms of  $Z(\eta)$, namely
\beqa
Z_p(\eta)&=&\frac{1}{1-\eta}-\left(\frac{\langle\sigma\rangle\langle \sigma^3\rangle}{\langle \sigma^2\rangle^2}-1\right)\frac{3\eta}{(1-\eta)^2}\nn
&&+\frac{\langle \sigma^3\rangle^2}{\langle \sigma^2\rangle^3}\left[Z(\eta)-\frac{1}{1-\eta}\right].
\label{ZpHS}
\eeqa
Equation \eqref{ZpHS} allows one to obtain \emph{indirectly} the equation of state of the pure HS fluid for high densities beyond the freezing transition from the knowledge of the equation of state of a given mixture in the same density region.

\begin{figure}
\includegraphics[width=.9\columnwidth]{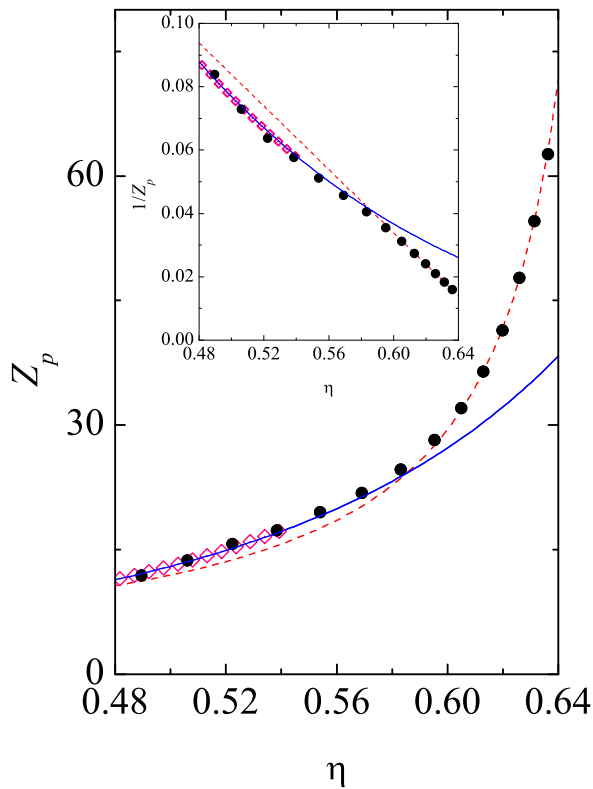}
\caption{Compressibility factor $Z_p(\eta)$ of the HS metastable fluid as a function of the packing fraction. Filled circles: results from Eq.\ \protect\eqref{ZpHS};  {open diamonds: simulation data from Ref.\ \protect\onlinecite{KLM04};} full line: compressibility factor corresponding to the CS equation of state; dashed line: compressibility factor arising from free volume considerations. The inset shows the same data in a different representation. \label{fig1}}
\end{figure}

Taking as input the {MC} data of Odriozola and Berthier\cite{OB11} for the pressure of the binary mixture  they studied ($x_1=x_2=\frac{1}{2}$, $\sigma_2/\sigma_1=\frac{7}{5}$), we first computed the corresponding compressibility factor $Z(\eta)$ and then substituted the values of this latter on the right-hand side of Eq.\ \eqref{ZpHS} to infer the compressibility factor $Z_p(\eta)$ of the pure HS fluid in the metastable fluid region. The results appear as full circles in Fig.\ \ref{fig1}, where we have also included {the accurate molecular dynamics (MD) simulation data reported by Kolafa et al.\cite{KLM04} in the metastable region. Figure \ref{fig1} incorporates as well the curves}  corresponding to the CS compressibility factor and  to a free volume compressibility factor of the form $Z_{\text{fv}}(\eta)\equiv {d}/{(1 - \eta/\eta_c)}$, with $d=3$ and where $\eta_c\simeq 0.668$ is the only fitting parameter. The inset contains the same data and curves, this time in the representation $1/Z_p(\eta)$ vs $\eta$. It is interesting to point out {the high degree of consistency in the region $0.48\leq\eta\leq 0.54$ between the MD simulation data for the true monodisperse HS system\cite{KLM04} and the values inferred via Eq.\ \eqref{ZpHS} from the MC simulation data for the binary mixture reported in Ref.\ \onlinecite{OB11}. Furthermore,} assuming that the present results for $Z_p(\eta)$ are correct {for $\eta>0.54$, we observe that} the CS compressibility factor remains accurate in the metastable fluid region up to $\eta\simeq 0.58$. On the other hand, beyond $\eta\simeq 0.58$ it is the free volume fit that gives an adequate description. We note that Odriozola and Berthier\cite{OB11} fitted the compressibility factor of the \emph{mixture} in the region $0.58\leq \eta \leq 0.63$ also with a free volume form, except that they obtained $\eta_c=0.669$ and adjusted the prefactor to be $d'=2.82$ instead of $d=3$. In view of Eq.\ \eqref{Ze3}, we find that the ratio $d'/d= 0.94$ can be explained by the factor multiplying $Z_p$, namely ${\langle \sigma^2\rangle^3}/{\langle \sigma^3\rangle^2}\simeq 0.93$, so that both free volume forms are fully compatible.

\begin{figure}
\includegraphics[width=.9\columnwidth]{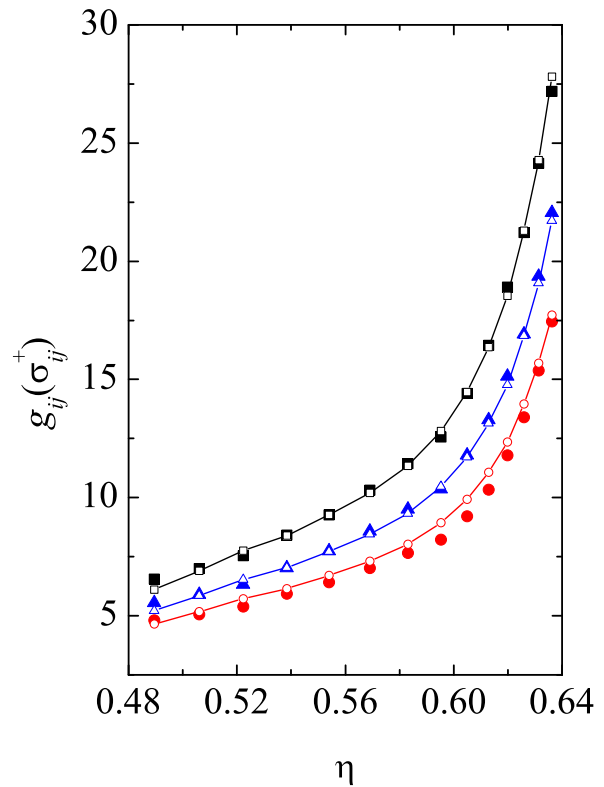}
\caption{Contact values at high density of the radial distribution functions of a binary mixture with $x_1=x_2=\frac{1}{2}$ and $\sigma_2/\sigma_1=\frac{7}{5}$. Open symbols, with solid lines drawn to guide the eye, are the results of Eq.\ \protect\eqref{5}, while the filled symbols are simulation data from Ref.\ \protect\onlinecite{OB11}: $g_{11}(\sigma_{1}^+)$ (red circles); $g_{22}(\sigma_{2}^+)$ (black  squares); $g_{12}(\sigma_{12}^+)$ (blue triangles).\label{fig2}}
\end{figure}

Once we have examined the possibility to infer the compressibility factor for a HS fluid in the metastable fluid region from the high density results of the compressibility factor of a binary mixture with $x_1=x_2=\frac{1}{2}$, $\sigma_2/\sigma_1=\frac{7}{5}$, we will check on the consistency of our approach by testing the validity of Eq.\ \eqref{5}. Since it is reasonable to expect that an accurate value of $g_{p}(\eta)$ should lead to accurate values of $g_{ij}(\sigma_{ij}^+)$, in order to assess the merits of our proposal we take for the contact value of the pure HS fluid the one that comes from the $Z_p(\eta)$ that we obtained from the use of Eq.\ \eqref{ZpHS}. The results for all pairs 11, 22, and 12 are shown in Fig.\ \ref{fig2}, where we have also included the simulation data of Odriozola and Berthier\cite{OB11} for these radial distribution functions for comparison. As clearly seen in this figure, the agreement between our computed values and those of simulation is very good. This lends support to the use of Eq.\ \eqref{5} also for high densities.

In summary, in this paper we have proposed a new route to derive the equation of state of a metastable HS fluid. In contrast with our previous work, which rested on the notion that the easier system was the pure HS fluid while the complicated one was the HS multicomponent mixture, here we have used the simulation results for a HS binary mixture at high density to infer the equation of state of a metastable {one-component} HS fluid. Although not shown, we have also checked that if, instead of using the simulation data for the pressure of the mixture, we use the simulation data for the contact values of the radial distribution functions $g_{11}(\sigma_{1}^+)$, $g_{22}(\sigma_{2}^+)$, and $g_{12}(\sigma_{12}^+)$ to compute the compressibility factor of the mixture $Z(\eta)$, the {results} that we get for $Z_p(\eta)$ {are} practically indistinguishable from the {ones} shown in Fig.\ \ref{fig1}.

On a different path, we have also {shown} that our recipe  to obtain contact values of the radial distribution functions of multicomponent HS mixtures from the contact value of the radial distribution function of the pure HS fluid also works for high densities. Given the fact that studying metastable HS fluids directly in simulations is very difficult, the present approach is offered as a {possible} useful alternative. Of course, one cannot reach definite conclusions concerning this alternative on the basis of the results of the single binary mixture analyzed in this paper. However the consistency in very many aspects of our approach as well as its relative simplicity are encouraging. We hope that the above serves to motivate more simulation studies like the one by Odriozola and Berthier\cite{OB11} for other mixtures at high density. Once they become available, the usefulness of our proposal may be further assessed.

\begin{acknowledgments}
We want to thank G. Odriozola for kindly providing us with the simulation data.
Two of us (A.S. and S.B.Y) acknowledge the financial support of the Ministerio de  Ciencia e Innovaci\'on (Spain) through Grant No. FIS2010-16587 and  the Junta de Extremadura (Spain) through Grant No.\ GR10158 (partially financed by FEDER funds). The work of M.L.H. has been partially supported by DGAPA-UNAM under project IN -107010-2.
\end{acknowledgments}


\end{document}